# Application of atomic magnetometry in magnetic particle detection

S. Xu, M. H. Donaldson, and A. Pines[a]

*Department of Chemistry, University of California at Berkeley, and Materials Sciences Division, Lawrence Berkeley National Laboratory, Berkeley, CA 94720*

S. M. Rochester, and D. Budker[b]

*Department of Physics, University of California at Berkeley, and Nuclear Sciences Division, Lawrence Berkeley National Laboratory Berkeley, CA 94720*

V. V. Yashchuk

*Advanced Light Source, Lawrence Berkeley National Laboratory, Berkeley, CA 94720*

We demonstrate the detection of magnetic particles carried by water in a continuous flow using an atomic magnetic gradiometer. Studies on three types of magnetic particles are presented: a single cobalt particle (diameter ~150 μm, multi-domain), a suspension of superparamagnetic magnetite particles (diameter ~1 μm), and ferromagnetic cobalt nanoparticles (diameter ~10 nm, 120 kA/m magnetization). Estimated detection limits are 20 μm diameter for a single cobalt particle at a water flow rate 30 ml/min, $5 \times 10^3$ magnetite particles at 160 ml/min, and 50 pl for the specific ferromagnetic fluid at 130 ml/min. Possible applications of our method are discussed.

[a]Electronic mail: pines@berkeley.edu

[b]Electronic mail: budker@berkeley.edu

Magnetic particles of micrometer and nanometer sizes are widely used in biomolecular labeling and cell separation[1-5], allowing manipulation of the components that are associated with the magnetic particles by an external magnetic field. These particles are also prevalent as contrast agents for magnetic resonance imaging[1-5].

In order to characterize the magnetization of these particles and monitor their behavior, a sensitive detection method is required. Several techniques have been developed for detecting weak magnetic fields, for example, superconducting quantum interference devices (SQUID)[6,7], giant magnetoresistive (GMR) sensors[8,9], magnetic resonance imaging (MRI)[10], vibrating sample magnetometers[11,12], and atomic magnetometers[13]. Each method has both advantages and disadvantages. For example, SQUIDs offer ultrahigh sensitivity and have been used extensively to detect weak magnetic signals, but they require cryogenics. GMR sensors are relatively convenient to use, however they require the sample to be extremely close (on the order of microns) to the sensors. While MRI is a powerful tool for noninvasive diagnostics, the cost of MRI machines severely limits their accessibility. Vibrating sample magnetometry has relatively low sensitivity.

Here we explore the application of atomic magnetometry to detecting magnetic particles. Atomic magnetometry has reached sensitivity comparable to that of SQUIDs[14,15] without requiring cryogenics. Details of our approach to atomic magnetometry are provided elsewhere[16]. Briefly, the magnetometer is based on nonlinear (in light power) magneto-optical rotation (NMOR) of laser light interacting with rubidium atoms contained in an anti-relaxation coated vapor cell. The frequency of the laser light is modulated (FM), and resonances in optical rotation are observed at modulation frequencies related to the Larmor precession frequency of the Rb atoms. The relationship between the external magnetic field B and the resonance modulation frequency ω is

$$\omega_M \approx 2g\mu(B_{bias} + B_{sample}),$$

where g is the atomic gyromagnetic ratio and μ is the Bohr magneton. A resonance occurs when the laser-modulation frequency is twice the Larmor precession frequency of the atoms. $B_{bias}$ is an applied magnetic field that is much greater than the sample field, $B_{sample}$, and so defines the detection axis. Therefore, the magnetic field from the sample along the direction of the bias field can be deduced from the frequency change of a magneto-optical resonance.

A schematic of our set-up is shown in Fig. 1. Two identical anti-relaxation coated $^{87}$Rb vapor cells inside a multi-layer magnetic shield form a first-order gradiometer that is insensitive to common-mode noise from environmental fluctuations. A long piercing solenoid generates a 0.5 G leading field ($B_{lead}$) which gives an orientation to the spins in the sample. Because of the geometry of the arrangement, the leading field is not "seen" by the magnetometer cells. A bias field of 0.7 mG ($B_{bias}$) gives an FM NMOR resonance frequency of ~1 kHz in the absence of the sample. When a magnetic sample is introduced to the detection region, it produces magnetic fields of opposite directions in the two cells. The signal from one arm of the gradiometer is continuously fed back to the laser modulation to keep this magnetometer on resonance. Thus the signal from the other magnetometer represents the difference field between the two cells created by the sample. We have achieved ~1 nG/Hz$^{1/2}$ sensitivity for near-DC signal[17] (for frequencies ~0.1 Hz), with 1-cm-sized cells separated by 1.5 cm.

We first measured the magnetization of a cobalt particle with an estimated diameter of 150 μm. The sample was embedded into a small piece of Styrofoam. Water carrying the foam flowed by a peristaltic pump through tubing (0.32 cm diameter) to the detection region of the gradiometer. As a control, an identical piece of Styrofoam without sample was also introduced into circulation. Figure 2 shows the results for two flow rates, 30 ml/min and 150 ml/min, which correspond to residence times of 30 ms and 160 ms, respectively, in the detection region. Each time the Styrofoam with the magnetic particle passed the gradiometer, a spike-like signal was produced, while the control Styrofoam produced no discernible signal. The average signal amplitude was much smaller for faster flow, since particle spent less time in the detection region. The magnitude and time dependence of the signal fluctuated between successive detections, most likely due to the random position and orientation of the particle in the detection region. From the signal-to-noise ratio in the slower flow, we estimate the detection limit to be a single cobalt particle with ~20 μm diameter. This estimation assumes multi-domain structure of the particles, and the scaling of their magnetic moment as square root of the volume. For single-domain particles, much smaller ones can be detected. In this case, we can estimate the detection limit to be ~5 μm diameter, given the present sensitivity of the gradiometer. The throughput can be increased up to 1200 ml/min using larger-diameter tubing, with the current spacing of 1 cm between the two cells. Therefore, such magnetic particles can be detected at essentially arbitrarily low concentrations in a large volume, and with high throughput.

Two types of smaller particles were measured similarly. One type was a superparamagnetic suspension containing amine-coated magnetite particles with ~1 μm diameter (Sigma-Aldrich, I7643). The sample was prepared by loading 18 nl of a suspension into a piece of capillary (diameter 150 µm, 1 mm length) and wrapping the capillary with Styrofoam. The total number of particles in the sample was ~4.5x$10^5$. The results are shown in Fig. 3, with water flow rate 160 ml/min. Panel (a) shows typical real time detection as the particles circulate. In order to measure the possible relaxation of the magnetization of the superparamagnetic particles, we continuously monitored the signal for over 1400 seconds. Averages of ten consecutive measurements are plotted versus the average measurement time after initial magnetization of the sample by a 3 kG permanent magnet. [Fig. 3(b)]. (To ensure full magnetization was reached, we also tried to use 20 kG field for magnetization, which made no substantial difference on the signal amplitude.) No significant decay was observed for the time span of the experiment. From the amplitude of the averaged signal, we obtained the current detection limit to be 0.2 nl, or 5x$10^3$ particles. The leading field was also varied. We observed no signal dependence on the magnitude of the leading field.

The other sample was a ferromagnetic fluid (Strem Chemicals, 27-0001, 120 kA/m magnetization) incorporating cobalt nanoparticles with diameter ~10 nm. The sample was loaded in a similar fashion to the superparamagnetic particles mentioned above. The ferromagnetic fluid with cobalt nanoparticles produced strong signal because of their high magnetization (Fig. 4, water flowing at 130 ml/min). From the average signal-to-noise ratio of 360, we estimated the smallest detectable amount to be 50 pl for this specific ferromagnetic fluid, with a detection time constant of 30 ms.

These experiments suggest diverse applications for our method. The ability to detect rare events (single particles) in a large amount of sample could be used for security applications to screen for magnetically labeled viruses in dilute environments or for in-line quality control devices for industrial processes involving magnetic products or impurities (for example, detection of ferromagnetic particulates in engine oil). Our method also has potential applications in biological and medical research. The ultrahigh sensitivity could allow detection of trace amounts of proteins, DNA, or antibodies that have been labeled by magnetic beads, and in the study of biochemical events associated with the aggregation of magnetic particles.

The detection limit could be improved significantly by further optimization and modification of the apparatus. For example, sensitivity can be improved by using additional sensor cells. A higher-order gradiometer can thus be formed, which could allow one to eliminate the need for magnetic shielding. Smaller alkali vapor cells[18] will also be investigated which can be put closer to the sample, improving the filling factor of the sample which, consequently enhances the detection limit, and allowing the method to be coupled with microfluidic applications

Acknowledgements and dedication: D. B. acknowledges inspiration for this work from Vladimir G. Budker, and wishes to dedicate it to his memory. This work was supported by the Director, Office of Science, Office of Basic Sciences, Materials Sciences Division of the U.S. Department of Energy, and by an ONR MURI grant.

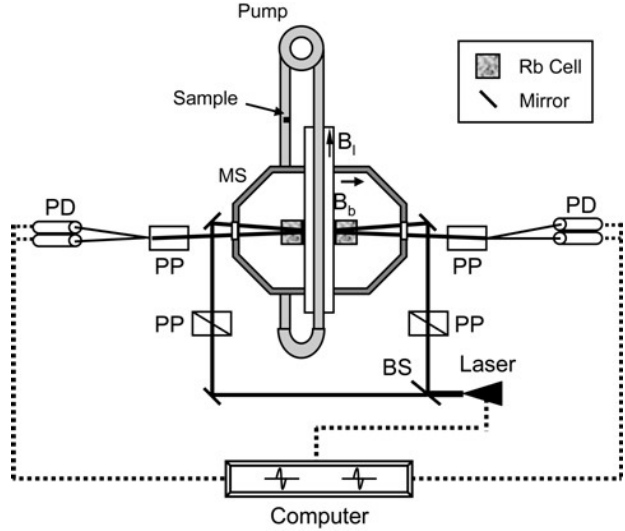

FIG. 1. Schematic of the setup for particle detection. BS: beam splitter; PP: polarizer prism; PD: photodiode; MS: magnetic shield.

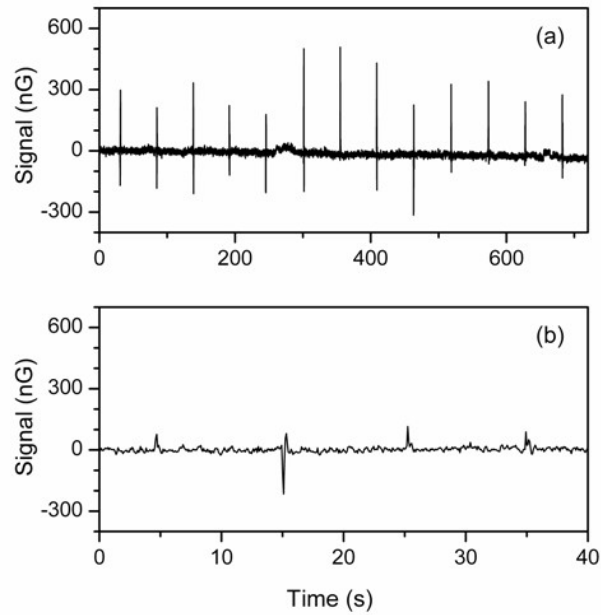

FIG. 2. Detection of a circulating cobalt particle carried by water at two different flow rates: (a) 30 ml/min; (b) 150 ml/min.

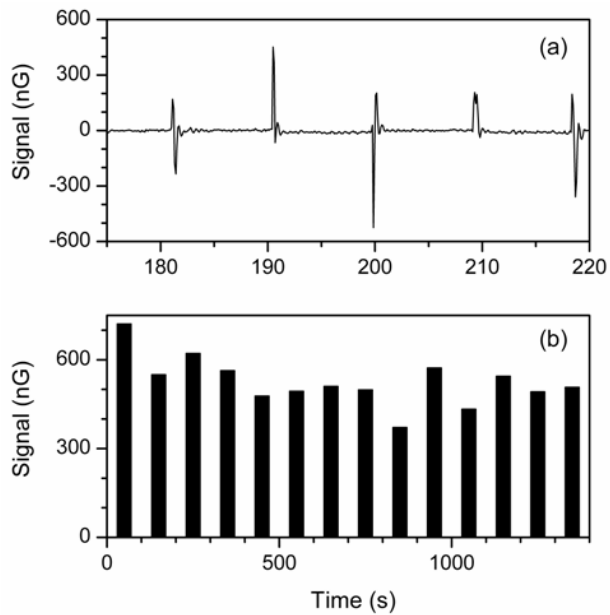

FIG. 3. Detection of 18 nl superparamagnetic magnetite suspension ($4 \times 10^5$ particles): (a) Typical real time detection; (b) Averaged signal (peak-to-peak) of ten consecutive measurements as a function of experimental time.

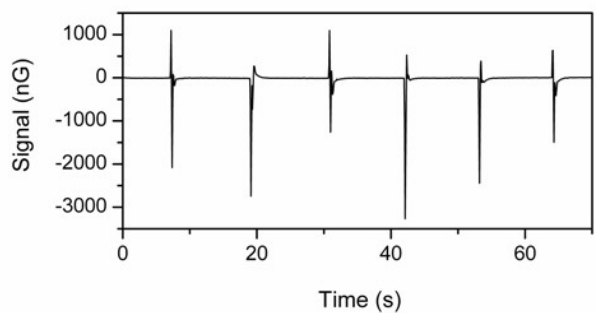

FIG. 4. Detection of 18 nl ferromagnetic fluid of cobalt nanoparticles (8.2% in kerosene, 1.5 kG magnetization).